\documentclass[a4paper]{article}

\usepackage{icrc2013}
\usepackage[english]{babel}

\title{The role of fast magnetic reconnection in acceleration zones of microquasars and AGNs}

\shorttitle{Fast magnetic reconnection in microquasars and AGNs}

\authors{
L.H.S. Kadowaki$^{1}$,
E.M. de Gouveia Dal Pino$^{1}$,
}

\afiliations{
$^1$  Universidade de S\~{a}o Paulo, IAG 
}

\email{lkadowaki@astro.iag.usp.br}

\abstract{Fast magnetic reconnection events, like those associated to solar flares, can be a very powerful mechanism operating at the jet launching region in the core of microquasars and AGNs. We have recently found that the magnetic power released by reconnection between the magnetic field lines of the coronal inner disk region and the lines anchored into the black hole is able to accelerate relativistic particles through a first-order Fermi process and produce the observed radio luminosity from both microquasars and low luminous AGNs. We also found that the observed correlation between the radio
luminosity and the mass of these sources, spanning $10^{9}$ orders of magnitude in mass,
is naturally explained by this process. Furthermore, recent observations of $\gamma$-ray emission with strong variability from AGNs and galactic compact sources indicate extremely small emission regions. In this work, assuming that this high energy emission is probably originated in the same acceleration zones that produce the radio emission, we have applied the scenario above of magnetic reconnection-driven acceleration in the magnetized corona around the accretion disks to investigate the origin of the high energy outcomes from an extensive number of sources including high and low luminous AGNs, microquasars, GRBs and the Crab nebula. As in the earlier analysis, we find that the radio emission of all microquasars and low luminous AGNs are well explained by our model. On the other hand, we find correlation of our model with the $\gamma$-ray emission only for microquasars and a few low luminous AGNs (e.g., M$87$), while none of the high luminous AGNs or GRBs are fitted, neither in radio (as previously found) nor in $\gamma$ emission. We attribute the lack of correlation of the $\gamma$ emission for most of the low luminous AGNs to the fact that this processed emission does not depend only on the local magnetic field activity around the source accretion disk (as the radio Synchrotron emission), but also on other environmental factors like the photon and density fields. We also find that the emission from both the high luminous AGNs and the GRBs actually anti-correlates with our model indicating that their emission is processed further out in the jet (as we claimed in \cite{dgdp_etal_10}, and in consistency with recent findings in \cite{nemmen_etal_12}). We conclude from these results that the emission we see from the low luminous AGNs and microquasars comes from the nuclear region of their sources and therefore, can be driven by nuclear magnetic activity connected to the source/accretion disk corona, as in our model. However, in the case of the high luminous AGNs (e.g. BL Lacs) and GRBs, the nuclear emission is blocked by the surrounding density and photon fields and, therefore, we can only see the jet emission further out (which has already lost correlation with the nuclear conditions). This also explains why high luminous AGNs do not fit the fundamental plane of luminosity source versus mass correlation.}

\begin{document}
\maketitle

\section{Introduction}

Galactic and extragalactic objects such as microquasars and AGNs often exhibit variability and quasi-periodic ejections of matter that may offer important clues about the physical processes that occur in their inner regions. 

A model to explain the origin of these ejections and the associated Synchrotron flare radio emission, which has characteristics that resemble those of solar flares, was developed by de Gouveia Dal Pino and Lazarian (\cite{dgdp_lazarian_05}; see also \cite{dgdp_etal_10}) where they invoked a process of fast magnetic reconnection between the magnetic field lines that arise from the accretion disk and the lines of the magnetosphere of the central source. In accretion episodes where the ratio between the effective disk pressure and magnetic pressure decreases to values smaller than the unity and the accretion rate approaches the critical Eddington rate, the magnetic reconnection may become very fast and release large amounts of magnetic energy power. Part of this energy heats the coronal and the disk gas and part accelerates particles to relativistic velocities through a first-order Fermi-like process 
(see also \cite{2011ApJ...735..102K}, \cite{2012PhRvL.108x1102K}, 
\cite{2013arXiv1302.4374D}) that results in a Synchrotron radio power-law spectrum  compatible with the observations. 
Using this model,  \cite{dgdp_lazarian_05} and \cite{dgdp_etal_10} also found that the observed correlation between the radio
luminosity and the mass from microquasars to low luminous AGNs, spanning $10^{9}$ orders of magnitude in mass,
is naturally explained by this process.

In recent years,  the very high energy emission (VHE) from AGNs have also revealed strong variability, with timescales of the order of days (e.g., M$87$), which points to extremely compact emission regions (corresponding to only a few Schwarzschild radii; e.g. \cite{abramowski_etal_12}). However, the localization of the emitting zones for several sources is until now unclear. Magnetic reconnection events occurring close to the black holes (BH) could offer appropriate conditions for producing particle acceleration and the associated VHE $\gamma$-ray emission in these sources. Similarly, microquasars are also expected to emit high-energy $\gamma$-rays owing to their general similarities to quasars (e.g., \cite{romero_etal_07}). Until now, only one source of this type, namely Cyg X-$3$, has been unambiguously detected in the GeV $\gamma$-rays, by the Agile and Fermi observatories (\cite{tavani_etal_09}, \cite{abdo_etal_09}). At TeV energies, only the flux upper limits are available, in spite of intensive monitoring. There is also some evidence of sporadic GeV-TeV $\gamma$-ray emission from another source of this type, Cyg X-$1$.

In the framework of high luminous AGNs and GRBs, Nemen et al. (\cite{nemmen_etal_12}) have recently shown that the jets produced by both classes of sources exhibit nearly the same correlation between the kinetic power carried by accelerated particles and the $\gamma$-ray luminosity. This is an indication that the bulk of the high-energy emission  comes from the jet region in these sources. On the other hand, the lack of correlation with low luminous AGNs may be an indication that the acceleration zones responsible for the high (and also low) energy emission in these cases are mostly in the core region, rather than further out in the jet, just like in the model  proposed above. Since the $\gamma$-ray emission is probably originated in the same acceleration zones that produce the radio emission, in the present work we  apply the de Gouveia Dal Pino et al. scenario above of reconnection-driven  acceleration in the magnetized corona of the accretion disks (\cite{dgdp_lazarian_05} and \cite{dgdp_etal_10}) and investigate if the high energy outcomes of microquasars and low luminous AGNs can be also interpreted in the light of this mechanism.
 
\section{Rate of magnetic energy released by magnetic reconnection: application to radio emission}

To evaluate the amount of magnetic energy that can be extracted through violent magnetic reconnection, it is adopted the standard model for the radiation-dominated accretion disk by \cite{shakura_sunyaev_73} and the model by \cite{liu_etal_02} to quantify the parameters of the corona. Also, it is assumed that the inner radius of the accretion disk corresponds approximately to the last stable orbit around the BH ($R_{X} = 3R_{S}$, where $R_{S}$ is the Schwarzschild radius). To determine the accretion rate immediately before an event of violent magnetic reconnection, it is assumed the equilibrium between the disk gas ram pressure and the magnetic pressure of the magnetosphere anchored at the BH horizon. It is assumed further that the intensity of the BH horizon field is of the order of that of the inner disk. Under these conditions one can show that the magnetic energy power released during violent magnetic reconnection for microquasars is approximately given by \cite{dgdp_etal_10}:

 \begin{equation}
  \dot{W}_{B} \simeq 10^{35} \alpha_{0.5}^{-19/16} \beta_{0.8}^{-9/16} M_{14}^{1/2} l_{100R_{X}}^{11/16}~~erg/s,
 \end{equation}
and for AGNs by \cite{dgdp_etal_10}:
 \begin{equation}
  \dot{W}_{B} \simeq 10^{40} \alpha_{0.5}^{-19/16} \beta_{0.8}^{-9/16} M_{8}^{1/2} l_{10^{4}R_{X}}^{11/16}~~erg/s,
 \end{equation}
where $\alpha_{0.5} = \alpha/0.5$ is the disk viscosity and $\beta_{0.8} = \beta/0.8$ is defined as the ratio between the effective disk pressure and the magnetic pressure. The BH mass $M_{14} = M/14M_{\odot}$ and $M_{8} = M/10^{8}M_{\odot}$ are parameters suitable for the microquasars and AGNs, respectively, and the parameters $l_{100R_{X}} = l/100R_{X}$ and $l_{10^{4}R_{X}} = l/10^{4}R_{X}$ are the scale height of the magnetic reconnection zone in the corona.
 
\begin{figure}[t]
  \centering
  \includegraphics[width=0.4\textwidth] {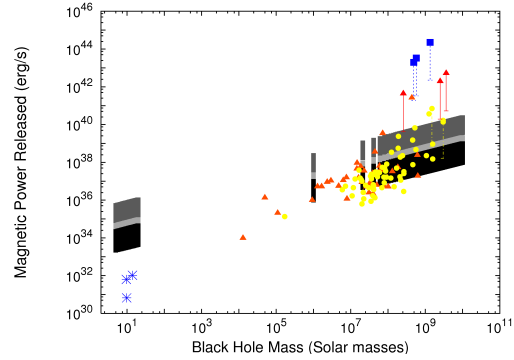}
  \caption{$\dot{W}_{B}$ versus the BH mass for both microquasars and AGNs. The stars represent the observed radio luminosities for microquasars. The circles, triangles and squares are observed radio luminosities of jets at parsec scales from LINERS, Seyfert galaxies, and luminous AGNs, respectively. The thick bars correspond to the calculated magnetic reconnection power and
encompass the parameter space that spans $5 M_{\odot} \leq M \leq 10^{10} M_{\odot}$, $0.05 \leq \alpha \leq 0.5$, $0.1 \leq \beta \leq 1$, and
$1 R_S \leq l \leq 1000 R_S$ (or $0.3 R_X \leq l \leq 333 R_X$), with $1 R_S \leq l \leq 10 R_S$ in black, $10 R_S < l \leq 30 R_S$ in light gray, and $30 R_S < l \leq 1000 R_S$ in dark gray. Extracted from \cite{dgdp_etal_10}.}
  \label{fig:dgdp_etal_10}
 \end{figure}

Figure \ref{fig:dgdp_etal_10} extracted from \cite{dgdp_etal_10} depicts a synthesis of the magnetic reconnection scenario for relativistic sources including both microquasars and AGNs. The diagram shows the calculated magnetic power released in violent magnetic reconnection events as a function of the central source mass for a suitable choice of the parameter space. The symbols correspond to the observed radio luminosities of superluminal components (stars for microquasars, circles and triangles for the low luminous AGNs, i.e. LINERs and Seyfert galaxies, respectively, and squares for luminous AGNs).

The diagram indicates that the magnetic power released during violent reconnection events obeys a correlation that is maintained throughout this interval, spanning $10^{9}$ orders of magnitude. This correlation implies an almost linear dependence (in a log-log diagram), which is approximately independent of the physical properties of the accretion disks of these sources. Moreover, it is compatible with the so-called ``fundamental plane'' obtained empirically, which correlates the radio and X-rays emission of microquasars and AGNs with the masses of their BH (see \cite{merloni_etal_03}). Thus, the model of \cite{dgdp_lazarian_05} provides a simple physical interpretation for the existence of this empirical correlation as due to coronal magnetic activity in these sources. The diagram also reveals that the more luminous AGNs do not  obey the same correlation, possibly because the density around the coronal region in these sources is so high that it ``masks'' the emission due to the magnetic activity. The radio emission in these cases is possibly due to regions further out at the supersonic jet, where it has already expanded enough to become optically thin and visible and where the relativistic electrons are probably accelerated in shocks (see \cite{dgdp_etal_10}). This conclusion in de Gouveia Dal Pino et al. \cite{dgdp_etal_10} is compatible with the recent correlation found by Nemem et al. (\cite{nemmen_etal_12}). 

\section{The same scenario for the $\gamma$-ray emission}

Employing the model described above of \cite{dgdp_etal_10}, we selected a preliminary set of compact sources including high and low luminous AGNs, GRBs and galactic sources (e.g., the pulsar Crab and the microquasars Cyg-X$1$ and Cyg-X$3$) with observed $\gamma$ and radio emission and compared their luminosities with the calculated magnetic reconnection power as a function of their mass.

  \begin{figure}[t]
  \centering
  \includegraphics[width=0.4\textwidth] {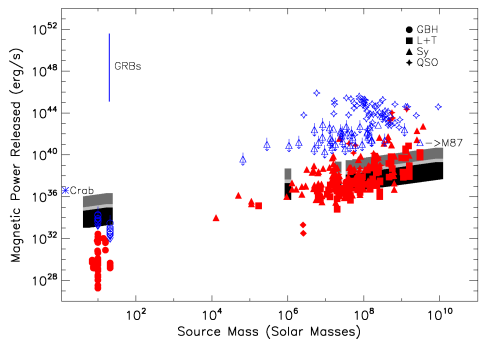}
  \caption{Magnetic power versus the central source mass. The open blue symbols correspond to the $\gamma$-ray emission from blazars, low luminous AGNs and galactic black holes (Cyg-X$1$ and Cyg-X$3$). The blue vertical line and blue star correspond, respectively, to the $\gamma$ emission from GRBs and the Crab pulsar. The filled red symbols correspond to the radio emission of different types of sources (Seyferts, liners, radio galaxies and microquasars).}
  \label{fig:icrc_13}
 \end{figure}

Figure \ref{fig:icrc_13} depicts the preliminary results of this study. As in Figure \ref{fig:dgdp_etal_10}, it compares the calculated magnetic power released by fast magnetic reconnection at the corona/accretion disk region as a function of the source mass with the observed $\gamma$ and radio luminosities. We have considered the $\gamma$ emission (open blue symbols in Figure \ref{fig:icrc_13}) of an extensive number of high luminous AGNs and GRBs (all obtained from \cite{nemmen_etal_12}), low luminous AGNs (Seyferts, liners and radio galaxies; see \cite{ackermann_etal_12}, \cite{aharonian_etal_03}, \cite{aharonian_etal_09} and \cite{aleksic_etal_2010}) and microquasars (Cyg-X$1$ \cite{albert_etal_07} and Cyg-X$3$ \cite{aleksic_etal_2010b}). In the case of GRBs (blue vertical line in the diagram), we have assumed that all sources have a mass around a few tens of solar masses. We have also included the radio emission counterparts of several of these sources (filled red symbols; obtained from \cite{merloni_etal_03}, and \cite{nagar_etal_05}) and the recently detected $\gamma$-ray emission from the Crab pulsar (blue star) which apparently comes from the surrounds of the central source (see \cite{tavani_etal_11}).


\section{Discussion and Conclusions}

As in the previous works (\cite{dgdp_lazarian_05} and \cite{dgdp_etal_10}), we clearly see in Figure \ref{fig:icrc_13} that the observed radio luminosity of the low luminous AGNs and microquasars can be explained by the magnetic power released by fast reconnection in the core region of these sources. This emission is due to Synchrotron radiation from relativistic electrons which can be accelerated by a first-order Fermi process directly within the magnetic reconnection site in the coronal region around the accretion disk (\cite{dgdp_lazarian_05}, \cite{2011ApJ...735..102K}, \cite{2012PhRvL.108x1102K}, \cite{2013arXiv1302.4374D}). The corresponding $\gamma$-ray emission from these sources, which is produced from the interaction of the accelerated relativistic electrons and protons with the surrounding photon and density fields (through inverse Compton, and/or pp inelastic collisions and photon-meson decay) can in principle be also associated with the same emission zone in the surroundings of the core of these sources. However,  direct  evidence for this association, though found for the microquasars and the Crab, is not found for most of the low luminous AGNs. This is explained by the fact that the $\gamma$ emission (contrary to the radio Synchrotron emission) does not depend only on the local magnetic fields, but also on the photon and density fields in the surroundings of the source/accretion disk, as stressed above, and these factors can provoke the loss of correlation with our nuclear emission model (induced by magnetic activity around the accretion disk).

The high luminous AGNs and the GRBs in Figure \ref{fig:icrc_13}, on the other hand, clearly do not have their radio or $\gamma$ emission correlated with the magnetic reconnection power released at the core regions. This result confirms the previous findings of \cite{dgdp_etal_10} and \cite{nemmen_etal_12} (see also \cite{2010IJMPD..19..729D}) which suggested that the $\gamma$ and radio emission observed in such sources is originated further out at the relativistic jet associated to these sources (as the nuclear emission is being screened by the surrounding strong photon and density fields).

 The results above connecting both the radio and $\gamma$ emission from low luminous compact sources to magnetically dominated reconnection processes in their nuclear regions, though preliminary are very promising as they suggest a unifying single process of relativistic particle acceleration in the core region of low luminous AGNs and compact galactic sources which naturally interpret the Fundamental Plane \cite{merloni_etal_03}. In forthcoming work, we intend to extend the analysis of the diagram above including more radio compact sources with $\gamma$-ray emission counterparts in order to reinforce the present conclusions.

\vspace*{0.5cm}
\footnotesize{{\bf Acknowledgment: This work has been partially supported by the Brazilian funding agencies FAPESP, CNPq and CAPES.}}

\end{document}